\documentclass{article}

\PassOptionsToPackage{numbers, compress}{natbib}
\usepackage[dblblindworkshop, final]{neurips_2026}

\workshoptitle{AI for Drug Discovery: Bridging the Translation Gap}
\usepackage[utf8]{inputenc}
\usepackage[T1]{fontenc}
\usepackage{hyperref}
\usepackage{url}
\usepackage{booktabs}
\usepackage{amsmath}
\usepackage{amssymb}
\usepackage{amsfonts}
\usepackage{amsthm}
\usepackage{nicefrac}
\usepackage{microtype}
\usepackage{xcolor}
\usepackage{graphicx}
\usepackage{caption}
\usepackage{enumitem}
\usepackage{svg}  

\newcommand{\bp}{\mathbf{p}}
\newcommand{\bq}{\mathbf{q}}
\newcommand{\bk}{\mathbf{k}}

\newcommand{\bz}{\mathbf{z}}
\newcommand{\bR}{\mathbf{R}}

\newcommand{\bA}{\mathbf{A}}
\newcommand{\bJ}{\mathbf{J}}
\newcommand{\bD}{\boldsymbol{\Delta}}
\newcommand{\btheta}{\boldsymbol{\theta}}
\newcommand{\Mask}{\mathcal{M}}
\newcommand{\Ody}{\textsc{Tessera}}

\title{Spatially Informed and Direction Sensitive Representation Learning for Spatial Transcriptomics}

\author{%
\begin{tabular}{c}
Azim Dehghani Amirabad$^{1}$\thanks{Equal contribution} \quad
Ali Khodabandeh Yalabadi$^{1}$\footnotemark[1] \quad
Artem Moskalev$^{2}$ \\[0.4em]
Yun-Ching Chen$^{1}$ \quad
Mehdi Pirooznia$^{1}$ \quad
Scott Oloff$^{1}$ \quad
Rui Liao$^{1}$\thanks{Corresponding author: Rui Liao (RLiao2@its.jnj.com)}
\end{tabular}
\\[0.8em]
$^{1}$Johnson \& Johnson Innovative Medicine \qquad
$^{2}$Independent Researcher
}

\begin{document}

\maketitle

\begin{abstract}
Cellular states are shaped by both intrinsic gene expression programs and the spatial microenvironment in which cells reside. Spatial transcriptomics makes this context measurable, but effective representation learning requires capturing both spatially informative molecular signals and intercellular organization. Existing approaches address these challenges only partially: conventional gene selection can overlook spatially structured signals, while methods for encoding cell--cell relationships often either reduce a cell's microenvironment to proximity alone, discarding direction, or rely on absolute coordinates tied to an arbitrary tissue frame. Consequently, cells with similar neighbor distances but distinct spatial organization may receive overly similar representations, despite experiencing different biological contexts that shape their cellular states. We introduce \textbf{Tessera}, a spatial representation learning framework that addresses both molecular selection and cellular context. Tessera combines spatially informed gene selection with a geometric inductive bias that contextualizes each cell within its local tissue environment. By modeling relative cell--cell relationships, Tessera is translation-invariant while preserving direction, allowing neighboring cells to contribute differently according to their spatial context and cellular state. Across held-out tissues and technologies, Tessera outperforms existing spatial foundation models in representation quality and generalizes zero-shot to contexts up to $16\times$ larger than those seen during training.
\end{abstract}

\section{Introduction}
\label{sec:intro}

Spatial transcriptomics (ST) measures gene expression while preserving the physical
organization of cells within tissue, enabling cellular states to be studied in their
native spatial context \cite{tian2023expanding,walker2022deciphering}. Because cellular
phenotypes are shaped by neighboring cells, local signaling, and tissue organization,
spatial context is an integral component of cellular state. This is particularly
relevant to drug discovery, where therapeutic targets, mechanisms of action, and
treatment response can depend on the tissue context in which molecular programs arise
\cite{pang2024drugtargets,cao2024spatialdrug}. Learning transferable representations
that jointly capture cellular state and spatial context could therefore support
context-aware target discovery, biomarker identification, and patient stratification
across tissues and disease settings \cite{aung2025spatial}.

Recent foundation models have demonstrated the value of large-scale representation
learning for single-cell transcriptomics \cite{cui2024scgpt,hao2024scfoundation},
motivating their extension to spatial omics
\cite{tejada2025nicheformer,zhao2025stofm,blampey2025novae}. Spatial data introduce
two complementary challenges for representation learning:
(i) selecting spatially informative genes to represent each cell, and
(ii) leveraging intercellular spatial relationships to capture cellular states within
their native tissue context. Existing gene-selection strategies capture only part of
the information needed for the first challenge. Highly expressed genes can be
dominated by broadly expressed housekeeping programs
\cite{stark2019rna,he2020integrating}, HVGs prioritize expression variability without
explicitly accounting for spatial organization \cite{stuart2019comprehensive}, and
SVGs emphasize tissue-level spatial patterns without considering the cell-specific
informativeness of each gene
\cite{sun2020statistical,chen2024evaluating,liang2024prost}.

The second challenge is to encode the spatial context that shapes cellular state.
A cell's microenvironment is inherently relational: neighboring cells influence its
state through their spatial organization relative to the cell of interest, rather than
through absolute slide coordinates. Existing approaches represent this context using
neighborhood or graph structure, pairwise distances, or absolute positional information
\cite{wen2023cellplm,zhao2025stofm,blampey2025novae}. Each captures only part of the
underlying tissue geometry: neighborhood-based encodings discretize continuous spatial
relationships, pairwise distance preserves proximity but discards direction, and
absolute coordinates retain direction but depend on an arbitrary tissue frame. Yet
tissue organization can be directionally structured
\cite{hu2021spagcn,blampey2025novae}, so equally distant cells may provide distinct
biological context. This motivates representing the \emph{full relative displacement}
between cells, preserving both distance and direction while remaining independent of
absolute position.


To address these complementary gaps, we introduce \textbf{Tessera}, a spatial
representation learning framework that jointly models spatially informative gene
expression and the relative organization of surrounding cells. For each cell, Tessera
constructs a spatially informed top gene set $k$ by combining cell specific expression
signals with tissue level spatial expression patterns, and encodes the resulting cell profile
with a pretrained scRNA-seq foundation model \cite{cui2024scgpt}. The resulting cell
embeddings are then contextualized through displacement-relative spatial attention,
instantiated with two-dimensional RoPE \cite{su2024roformer}. This imposes a geometric
inductive bias that preserves both distance and direction, allowing the model to
distinguish similarly distant cells that provide different biological contexts rather
than treating them as equivalent based on proximity alone. The resulting spatial representation is biologically grounded and highly
transferable, consistently improving expression imputation across held-out tissues and
technologies and generalizing zero-shot to spatial contexts up to $16\times$ larger
than those seen during training.

Our main contributions are:
(i) a \textbf{spatially informed gene-ranking strategy} that integrates
cell-level and tissue-level measures of spatial gene informativeness; and
(ii) \textbf{direction-sensitive spatial attention} that is translation-invariant
while preserving relative distance, direction, and physical spatial scale.

\begin{figure}[t] \centering 
\includegraphics[width=1\textwidth]{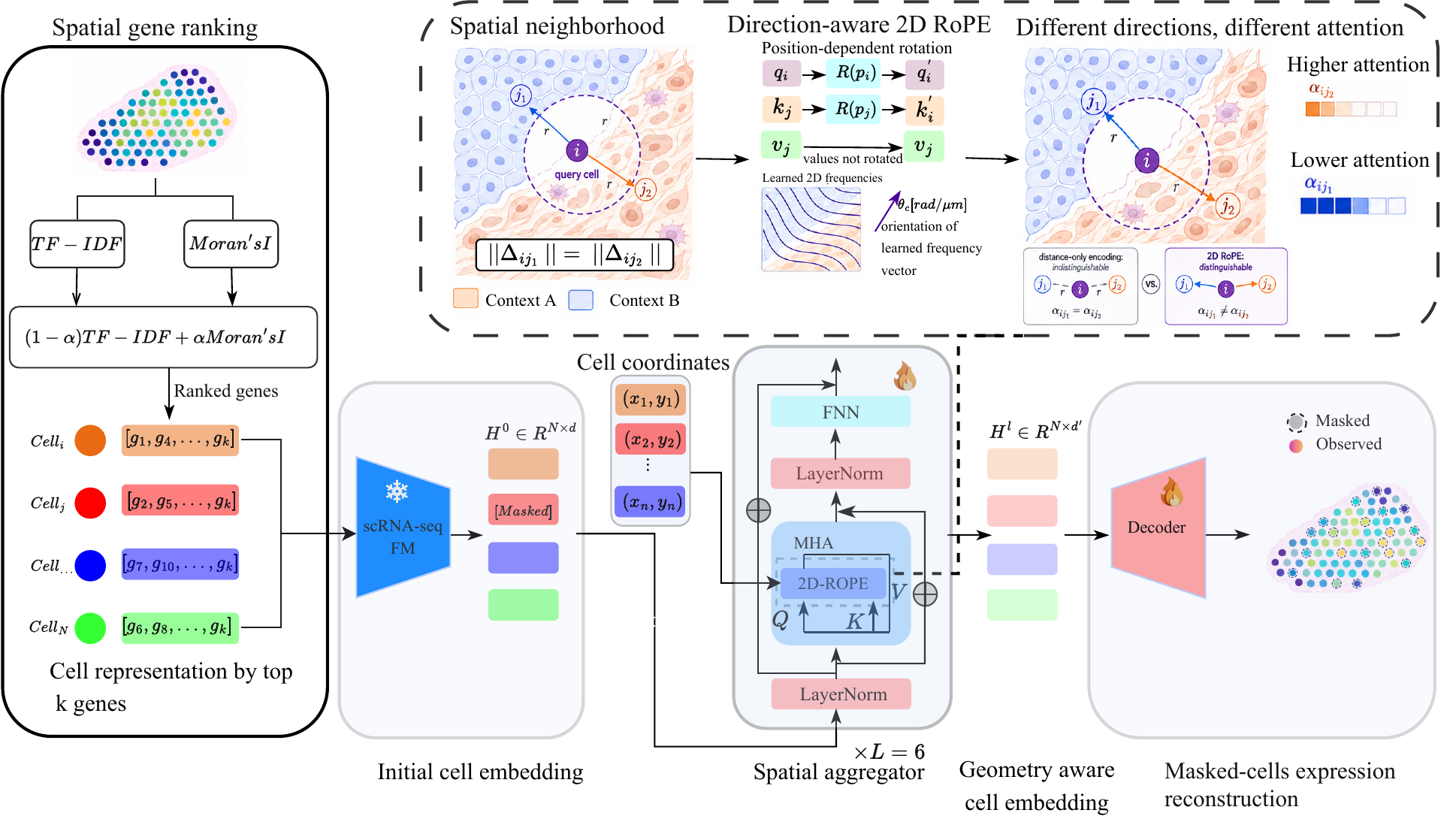}
\caption{\Ody\  learns spatially contextualized cell representations in two stages. Spatially informed ranking (SIR) selects spatially informative genes to represent cells, encoded by a pretrained scRNA-seq foundation model. The resulting cell embeddings are then contextualized by a geometry-aware aggregator with 2D-RoPE, letting neighbors contribute according to their relative distance and direction while remaining translation-invariant. Learning proceeds by masking cells and reconstructing masked cell expression from their spatial neighborhood.} 
\label{fig:arch} 
\vspace{-6mm}
\end{figure}



\section{Spatially informed gene ranking}

To construct informative gene sets for each cell, we introduce
\emph{spatially informed ranking} (SIR), which combines
cell-specific expression relevance with tissue-level spatial
structure. TF--IDF assigns each gene a cell-specific score
that favors genes enriched in that cell but less ubiquitous
across the section, while Moran's $I$ measures tissue-level
spatial autocorrelation. Because these statistics lie on
different scales, we combine their ranks rather than their
raw values to construct a spatially informed gene ranking
for each cell\cite{amirabad2026pastel}. The complete formulation and implementation
details are provided in Appendix~\ref{app:sir}.

\section{Direction-Sensitive Attention for Spatial Transcriptomics}
\label{sec:ropepope}

Let $\mathbf{c}_i$ denote the representation of cell $i$ and
$\mathbf{p}_i\in\mathbb{R}^2$ its physical position in micrometres. We seek to
contextualize each cell using the geometry of surrounding cells without tying the
representation to the arbitrary coordinate origin of a tissue section. Existing
approaches encode spatial context through absolute positions, pairwise distances, or
neighborhood graphs \cite{wen2023cellplm,zhao2025stofm,blampey2025novae}, but these
representations retain different aspects of the underlying geometry. We instead model
the full relative displacement
\[
    \Delta_{ij} = \mathbf{p}_j-\mathbf{p}_i ,
\]
between cells $i$ and $j$. This induces three desirable properties:

\begin{enumerate}[label=(\roman*),leftmargin=2.2em,itemsep=1pt,topsep=2pt]
    \item \textbf{Translation invariance:} attention depends on relative displacement,
    and is therefore independent of the tissue-section origin.
    
    \item \textbf{Direction sensitivity:} both the distance and bearing of neighboring
    cells are retained, allowing equally distant cells to provide distinct spatial
    context.
    
    \item \textbf{Physical-scale parameterization:} spatial relationships are expressed
    in physical units, providing a consistent notion of distance across technologies.
\end{enumerate}

We realize this geometric inductive bias using two-dimensional RoPE
\cite{su2024roformer}, adapting rotary positional encoding to continuous tissue
coordinates. Spatial frequencies are parameterized in radians per micrometre, so the
attention interaction depends on relative cell displacement while retaining both
distance and direction.

\paragraph{RoPE.} For each attention head of width $d_h$, we split the head into $d_h/2$ two-dimensional planes. Plane $c$ has a learnable frequency $\boldsymbol{\theta}_c\in\mathbb{R}^2$, defining the phase
\begin{equation}
    \psi_c(\mathbf{p})=\boldsymbol{\theta}_c^\top\mathbf{p}.
\end{equation}
Stacking the corresponding planar rotations gives the block-diagonal matrix
\begin{equation}
    \mathbf{R}(\mathbf{p})
    =
    \operatorname{diag}_{c}\,
    \mathbf{R}_2\!\left(\psi_c(\mathbf{p})\right),
\end{equation}
where
\begin{equation}
    \mathbf{R}_2(\psi)
    =
    \begin{pmatrix}
        \cos\psi & -\sin\psi\\
        \sin\psi & \cos\psi
    \end{pmatrix}.
\end{equation}
The attention logit for head $h$ is
\begin{equation}
    A_{ij}^{(h)}
    =
    \frac{
        \left\langle
        \mathbf{R}(\mathbf{p}_i)\mathbf{W}_Q^{(h)}\mathbf{c}_i,\,
        \mathbf{R}(\mathbf{p}_j)\mathbf{W}_K^{(h)}\mathbf{c}_j
        \right\rangle
    }{\sqrt{d_h}}
    +m_{ij},
    \qquad
    m_{ij}=
    \begin{cases}
        0, & \lVert\Delta_{ij}\rVert\le d_{\max},\\
        -\infty, & \text{otherwise},
    \end{cases}
    \label{eq:logit}
\end{equation}
where $\mathbf{c}_i\in\mathbb{R}^{d}$ is the cell token from the frozen encoder (Fig.~\ref{fig:arch}), $\mathbf{W}_Q^{(h)},\mathbf{W}_K^{(h)}\in\mathbb{R}^{d_h\times d}$ are the query and key projections, and $d_{\max}=200\,\mu\mathrm{m}$. The values are left unrotated, so the positional transformation affects how cell representations are compared rather than what is transported; $m_{ij}$ is a hard spatial cutoff rather than a learned function of distance.

Define
\begin{equation}
    \mathbf{q}_i^{(h)}=\mathbf{W}_Q^{(h)}\mathbf{c}_i,
    \qquad
    \mathbf{k}_j^{(h)}=\mathbf{W}_K^{(h)}\mathbf{c}_j.
\end{equation}
Because planar rotations compose additively and $\psi_c$ is linear,
\begin{equation}
    \left\langle
    \mathbf{R}(\mathbf{p}_i)\mathbf{q}_i^{(h)},\,
    \mathbf{R}(\mathbf{p}_j)\mathbf{k}_j^{(h)}
    \right\rangle
    =
    \mathbf{q}_i^{(h)\top}
    \mathbf{R}(\Delta_{ij})
    \mathbf{k}_j^{(h)}.
\end{equation}

Thus, $A_{ij}^{(h)}$ depends on position only through $\Delta_{ij}$, giving translation invariance by construction. Because $\boldsymbol{\theta}_c$ is a vector rather than a scalar, the relative phase $\boldsymbol{\theta}_c^\top\Delta_{ij}$ depends on the bearing of the displacement as well as its length, giving direction sensitivity. Finally, $\lVert\boldsymbol{\theta}_c\rVert=2\pi/\ell_c$ defines a wavelength $\ell_c$ in micrometres. We initialize $\{\ell_c\}$ on a geometric grid from $10\,\mu\mathrm{m}$ to $2000\,\mu\mathrm{m}$, with the bearings of $\boldsymbol{\theta}_c$ uniformly distributed on $[0,\pi)$.

\paragraph{PoPE.}
In spatial omics, cell identity and spatial location provide distinct signals, suggesting that their contributions to attention should be separable. However, RoPE may couple content and position within each query--key component ~\cite{gopalakrishnan2026pope}, which can make the positional phase depend on the content being matched. PoPE removes this coupling by representing each component with a non-negative magnitude and a phase determined solely by position~\cite{gopalakrishnan2026pope}. We therefore also adapt PoPE to continuous two-dimensional tissue coordinates. The resulting attention score takes the form
\begin{equation}
    A_{ij}^{\mathrm{PoPE}}
    =
    \sum_{c}
    \mu_{i,c}^{Q}\mu_{j,c}^{K}
    \cos\!\left(
        \boldsymbol{\theta}_c^\top\Delta_{ij}+\delta_c
    \right)
    + m_{ij},
\end{equation}
where $\mu_{i,c}^{Q},\mu_{j,c}^{K}\geq 0$ are the query and key magnitudes, $\boldsymbol{\theta}_c\in\mathbb{R}^2$ is a spatial frequency, and $\delta_c$ is a fixed phase offset. This preserves the displacement-relative, direction-sensitive geometry of RoPE while removing the content-dependent phase interaction, providing a controlled alternative under the same physical-scale parameterization.


\paragraph{Learning objective.}
We train Tessera with cross-cell masked reconstruction, holding out $10\%$ of cells
in each window by replacing their cell embeddings with a learned mask token while
retaining their coordinates. Because no expression-derived information from a masked
cell enters the spatial attention module, its profile must be inferred from neighboring
cells and their geometry. A gene-wise negative-binomial decoder reconstructs raw counts,
with loss computed only on masked cells; gradients update only the spatial aggregator
and decoder. Full masking, offset, and likelihood details are provided in
Appendix~\ref{app:objective}.

\section{Experiments}
\label{sec:exp}

We organize our experiments around the main design choices of Tessera. We first
evaluate spatially informed gene ranking (Appendix~\ref{app:ranking}), then isolate the contribution of spatial
geometry and compare positional encodings under a controlled architecture. We next
characterize the learned attention in terms of locality, direction sensitivity, and
invariance, and assess zero-shot extrapolation beyond the spatial scale seen during
training. Finally, we benchmark Tessera against existing spatial foundation models. Models, optimization, and the full experimental protocol are described in Appendix~\ref{app:experiments}.

\subsection{Impact of spatial geometry}
\label{sec:results}

\begin{table}[t]
\centering
\scriptsize
\caption{\textbf{Impact of spatial geometry.} Held-out expression-imputation
Pearson correlation across nine sections. $\Delta$ and $t$ are paired section-wise
comparisons with \Ody{}. The geometry-free ablation differs only by removing
positional information.}
\label{tab:main}
\begin{tabular}{lccc}
\toprule
Predictor & Pearson $r$ & $\Delta$ vs.\ \Ody{} & $t$ \\
\midrule
\Ody{} without geometry & $0.26565$ & $-0.18370$ & $-10.49$ \\
\texttt HVG-PCA & $0.29072$ & $-0.15863$ & $-8.22$ \\
\texttt scGPT (no aggregator) & $0.29985$ & $-0.14950$ & $-7.97$ \\
\texttt Smoothed PCA & $0.30282$ & $-0.14653$ & $-7.81$ \\
\texttt Neighbour average & $0.30806$ & $-0.14129$ & $-7.41$ \\
\midrule
\textbf{\Ody} & $\mathbf{0.44935}$ & --- & --- \\
\bottomrule
\end{tabular}
\end{table}

\Ody{} outperforms all non-geometric baselines on every held-out section
(Table~\ref{tab:main}), improving Pearson correlation by $+0.141$ over the strongest
baseline and by $+0.184$ over the otherwise identical geometry-free ablation.
The gain is consistent across seeds, and coordinate-permutation controls further
confirm that the improvement arises from spatial geometry
(Appendix~\ref{app:experiments}).

\subsection{Relative geometry drives performance}
\label{sec:pe}
\begin{figure}[h]
\centering
\includegraphics[width=0.88\textwidth]{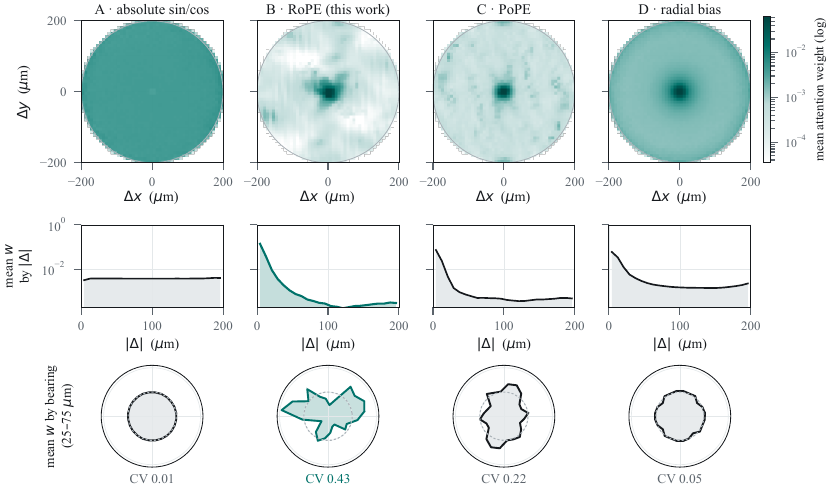}
\caption{\textbf{Direction sensitivity of learned spatial attention.} Attention patterns for four positional encodings on a held-out MERSCOPE
breast tissue section. \textbf{Top:} Mean attention weights as a function of relative cell
displacement, aggregated across eight attention heads using a shared
logarithmic scale.
\textbf{Middle:} Radial attention profiles showing spatial locality.
\textbf{Bottom:} Angular attention profiles within a fixed
25--75~$\mu$m distance range, normalized by their mean.
The coefficient of variation (CV) quantifies directional sensitivity,
with higher values indicating stronger anisotropy.
RoPE exhibits the strongest directional variation (CV = 0.43),
while the distance-only bias remains approximately isotropic
(CV = 0.05).}
\label{fig:f15}
\end{figure}

Holding the architecture and training protocol fixed, we vary only the positional encoding. Absolute sinusoidal encoding provides negligible improvement (+0.0005), whereas relative encodings improve Pearson correlation by +0.160–0.174 across all nine sections. RoPE performs best among the relative variants, although differences within this group are small and treated as descriptive. Beyond predictive performance, the learned attention patterns reveal distinct spatial behaviors (Fig.~\ref{fig:f15}). Relative encodings exhibit spatial locality, while RoPE shows the strongest directional variation within a fixed distance range, in contrast to the approximately isotropic distance-only bias. 
Full positional-encoding results and angular analyses are provided in Appendix~\ref{app:pe}, with per-section results in Table~\ref{tab:persection}.

\subsection{Invariance and spatial extrapolation}
\label{sec:invariance}

Directional encoding raises a natural question: tissue sections have no canonical
orientation, so why preserve direction rather than enforce rotation invariance? While
absolute orientation is arbitrary, relative direction can still carry biological
information because local tissue organization is often anisotropic: cells at the same
distance but different bearings may occupy distinct spatial contexts and therefore
provide different signals about cellular state. Translation invariance removes
dependence on the arbitrary section origin while preserving these relative directional
relationships. RoPE and PoPE therefore retain direction by design, whereas a
distance-only bias is isotropic and cannot distinguish equally distant cells at
different bearings (Appendix~\ref{app:pe}). Transformation tests confirm that RoPE and PoPE preserve attention under translation but change it under rotation, whereas the radial distance bias remains invariant to both (Fig.~\ref{fig:f16}; Table~\ref{tab:invariance}).

\begin{figure}[h]
\centering
\includegraphics[width=0.88\textwidth]{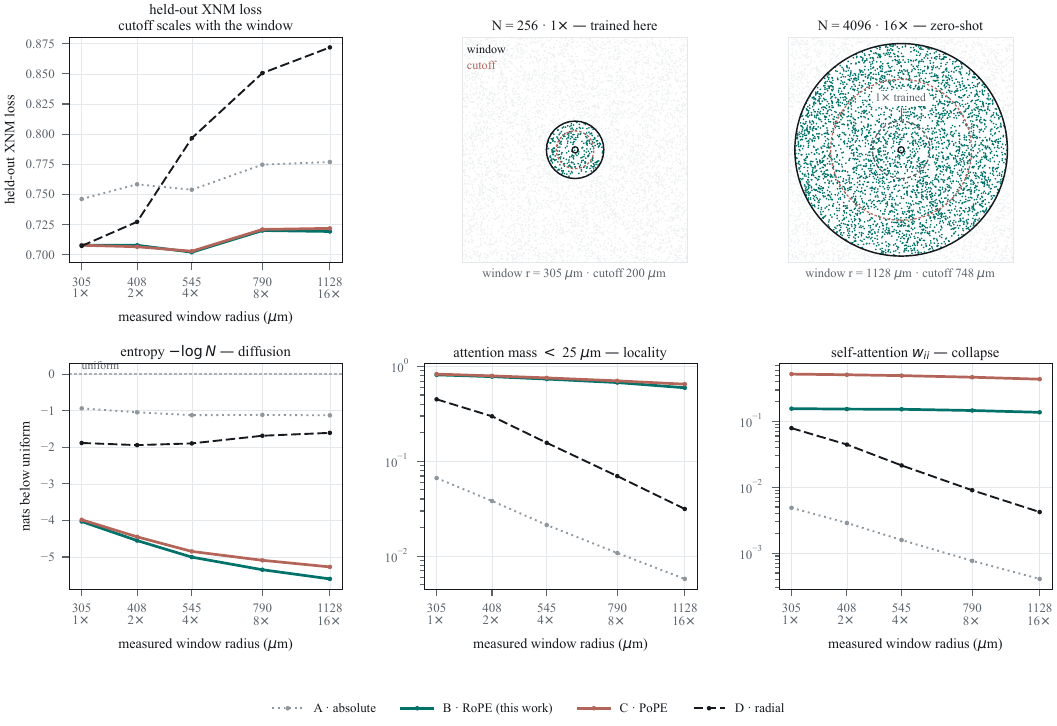}
\caption{\textbf{Zero-shot spatial extrapolation beyond the training window.}
Comparison of four positional encodings as the spatial context
expands from 256 ($1\times$) to 4096 ($16\times$) cells,
without retraining or positional rescaling.
\textbf{Top left:} Held-out cross-cell masked reconstruction
loss across increasing window sizes.
\textbf{Top right:} Illustration of the training and
extrapolation windows, showing the corresponding spatial
radii and scaled attention cutoffs.
\textbf{Bottom:} Attention entropy relative to $\log N$
(left), attention mass within 25~$\mu$m (middle),
and self-attention weight (right), characterizing
attention diffusion, locality, and collapse.
RoPE and PoPE exhibit limited performance degradation
(1.6\% and 2.0\%, respectively), compared with
23\% for the radial distance bias, demonstrating
stronger zero-shot extrapolation to larger spatial contexts.
}
\label{fig:f17}
\end{figure}

This relative formulation also supports spatial extrapolation.
Because displacement-based encodings are defined continuously in
physical space, they can be evaluated beyond the spatial scale
observed during training, whereas fixed-range distance representations
are tied to their predefined support. When evaluated zero-shot from
$N=256$ to $N=4096$ cells, RoPE and PoPE exhibit only 1.6\% and
2.0\% degradation, respectively, compared with 23\% for the
fixed-range distance bias (Fig.~\ref{fig:f17}).
The corresponding attention statistics further characterize how
locality and attention concentration evolve as the spatial context
expands. Additional analyses are provided in 
Appendix~\ref{app:extrap}.

\subsection{Comparison with spatial foundation models}
\label{sec:benchmark}

\begin{table}[t]
\centering
\scriptsize
\caption{\textbf{Comparison with spatial foundation models.}
Linear probes are trained on identical training-cell budgets and evaluated on the same
nine held-out sections. \emph{Left:} neighborhood-density prediction from individual
cell embeddings (Pearson $r$, 95\% CI). \emph{Right:} within- and cross-platform
transfer between Xenium (X) and MERSCOPE (M).}
\label{tab:benchmark}
\begin{tabular}{lcc@{\hskip 0.8em}cccc}
\toprule
& \multicolumn{2}{c}{Neighbourhood density} & \multicolumn{4}{c}{Transfer ($r$)} \\
\cmidrule(lr){2-3}\cmidrule(lr){4-7}
Model & $r$ & 95\% CI & X$\to$X & M$\to$M & M$\to$X & X$\to$M \\
\midrule
HVG$\to$PCA & $0.1452$ & $[0.0543,\,0.2414]$ & $0.187$ & $0.110$ & $-0.032$ & $0.034$ \\
Nicheformer & $0.2130$ & $[0.1037,\,0.3459]$ & $0.349$ & $0.173$ & $0.237$ & $0.004$ \\
scGPT-spatial & $0.2526$ & $[0.1150,\,0.4112]$ & $0.389$ & $0.209$ & $0.125$ & $0.037$ \\
\midrule
\textbf{\Ody} & $\mathbf{0.6246}$ & $\mathbf{[0.4897,\,0.7517]}$ & $\mathbf{0.694}$ & $\mathbf{0.555}$ & $\mathbf{0.540}$ & $\mathbf{0.409}$ \\
\bottomrule
\end{tabular}
\end{table}

\Ody{} substantially outperforms existing spatial foundation models under the same
linear-probe protocol, reaching $r=0.6246$ on neighborhood-density prediction versus
$0.2526$ for scGPT-spatial and $0.2130$ for Nicheformer (Table~\ref{tab:benchmark}).
This comparison is conservative, as Nicheformer was pretrained on the evaluation
sections whereas \Ody{} was not. The advantage also transfers across technologies:
training on Xenium and evaluating on MERSCOPE yields $r=0.409$ for \Ody{} versus
$0.004$ for Nicheformer, consistent with learning a transferable physical spatial
representation. Additional radius sweeps, seed analyses, and validation controls are reported in Appendix~\ref{app:benchmark}.

\section{Discussion and conclusion}
\label{sec:limits}

Tessera addresses two complementary challenges in spatial representation learning:
combining cell-specific expression relevance with tissue-level spatial informativeness
for gene selection, and contextualizing cellular states through intercellular spatial
relationships. Its geometric inductive bias reflects a simple property of tissue
organization: cellular context is relational, and equally distant cells can provide
different biological information depending on their relative position and cellular
state. The RoPE-based mechanism preserves this directional structure while remaining
invariant to arbitrary translations of the tissue coordinate system. Importantly, we
do not assume a universal tissue orientation; rather, we avoid imposing isotropy by
construction, allowing locally anisotropic tissue organization to be represented.
Although our current evaluation focuses on expression imputation and spatial probing,
such context-aware representations could ultimately support downstream applications
including spatial biomarker discovery, target prioritization, and prediction of
treatment response.

\bibliographystyle{plainnat}
\bibliography{odyssey}

\newpage
\appendix
\section{Related work}
\label{app:related}

The taxonomy of \S\ref{sec:intro} separates spatial models by whether the
continuous displacement between two cells ever reaches attention. It does not in
\emph{Nicheformer} \citep{schaar2025nicheformer}, which never consumes coordinates
during pretraining and relies on expression to carry the spatial trace. It is
collapsed to an unweighted mean in \emph{scGPT-spatial}
\citep{wang2025scgptspatial}, a choice made deliberately so that the model
transfers between sections. It enters absolutely in \emph{CellPLM}
\citep{wen2023cellplm}, as sinusoidal embeddings of $(x,y)$. It enters as an
isotropic scalar in \emph{SToFM} \citep{zhao2025stofm}, whose SE(2) Transformer is
adopted from the molecular model of \citet{zhou2023unimol} and whose distance
matrix is described as ensuring SE(2) invariance in a straightforward manner. And
it is binarised in the graph models
\citep{blampey2025novae,madhu2025heist,birk2025nichecompass}.

Rotary encodings were introduced for one-dimensional token sequences
\citep{su2024roformer}. Extending them beyond one dimension requires a
commuting-generator construction, developed for vision transformers by
\citet{heo2024rope}, generalised by \citet{ostmeier2024liere}, and given
exact-invariance guarantees for arbitrary-dimensional continuous coordinates by
\citet{schenck2025string}. PoPE \citep{gopalakrishnan2026pope} argues that the
rotary phase entangles content with position, because the two-dimensional
subcomponents of the query and key carry a content-dependent angle that adds to the
positional rotation, and removes it by applying a softplus so that every
subcomponent has a positive first coordinate and a zero second. Our contribution is
not the encoding but its application to physical tissue coordinates, the
physical-units parameterisation that makes learned scales cross-platform, the
controlled four-way comparison of \S\ref{sec:pe}, and the extrapolation result of
\S\ref{sec:invariance}.

Adding a learned function of pairwise distance to an attention logit originates
with structure-aware graph and molecular transformers
\citep{ying2021graphormer,zhou2023unimol,jumper2021alphafold,shuaibi2021spinconv}
\section{Proof of exact translation invariance}
\label{app:proof}

\begin{proof}
By the block-diagonal construction of the rotary encoding,
\begin{align}
\bR(\bp)^{\!\top}\bR(\bp')
&= \bigoplus_c
\bR_2(\psi_c(\bp))^{\!\top}
\bR_2(\psi_c(\bp')) \nonumber\\
&= \bigoplus_c
\bR_2\!\left(\psi_c(\bp')-\psi_c(\bp)\right).
\end{align}
Since the planar rotations compose additively and
$\psi_c(\bp)=\btheta_c^{\!\top}\bp$ is linear,
\[
\psi_c(\bp')-\psi_c(\bp)
= \psi_c(\bp'-\bp).
\]
Consequently,
\[
\bR(\bp)^{\!\top}\bR(\bp')
= \bR(\bp'-\bp).
\]
Substituting into the attention inner product in
Eq.~\eqref{eq:logit} gives
\begin{equation}
\left\langle
\bR(\bp_i)\bq_i,
\bR(\bp_j)\bk_j
\right\rangle
=
\bq_i^{\!\top}
\bR(\bD_{ij})\bk_j,
\qquad
\bD_{ij}=\bp_j-\bp_i.
\label{eq:relative_attention_proof}
\end{equation}
Thus, the attention logit depends on spatial coordinates
only through relative displacement, establishing exact
translation invariance.
\end{proof}

\paragraph{Numerical verification.}
We additionally verified the relative-phase identity
numerically in float64, obtaining a maximum absolute
deviation of $5.7\times10^{-14}$. The result remained
unchanged under a translation of
$(12.3,-8.9)\,\mathrm{mm}$, consistent with the
analytical result.

\paragraph{Extension to PoPE.}
The same translation-invariance property holds for
the two-dimensional adaptation of PoPE. With spatial
phase $\psi_c(\bp)=\btheta_c^{\!\top}\bp$, the
attention logit is
\begin{equation}
A_{ij}^{\mathrm{PoPE}}
=
\sum_c
\mu_{i,c}^{Q}\mu_{j,c}^{K}
\cos\left(
\btheta_c^{\!\top}\bD_{ij}+\delta_c
\right)
+m_{ij}.
\end{equation}
Here, the query and key magnitudes are independent
of spatial coordinates, and the fixed phase offset
$\delta_c$ does not depend on absolute position.
Assuming the spatial mask $m_{ij}$ depends only
on relative displacement, the complete attention
logit is therefore invariant to arbitrary
translations of the coordinate system.
This property is also verified empirically in
Table~\ref{tab:invariance}.

\paragraph{Frequency initialization and complexity.}
$\{\lVert\btheta_c\rVert\}$ is set on a geometric grid so that the wavelengths
$\ell_c = 2\pi/\lVert\btheta_c\rVert$ span $10\,\mu$m to $2000\,\mu$m, with
bearings spread uniformly on $[0,\pi)$. Frequencies carry a $0.1\times$
learning-rate multiplier, which we found necessary to prevent an early collapse of
attention entropy. Per window and layer, inter-cell attention costs $O(N^2 d)$: at
$N=256$ that is $6.6\times10^{4}$ pairs, negligible beside the backbone's
$O(NK^2 d)$ gene-level computation at $4.2\times10^{6}$ pairs, which is paid once
and cached because the backbone is frozen. The trained aggregator is $L=6$ layers,
$8$ heads, $d=512$, about $19$M parameters.

\paragraph{Why the construction generalises beyond one dimension.}
Equivalently $\bR(\bp) = \exp(p^x \bA_x + p^y \bA_y)$ with
$\bA_x = \bigoplus_c \theta^x_c \bJ$, $\bA_y = \bigoplus_c \theta^y_c \bJ$ and
$\bJ = \left(\begin{smallmatrix}0 & -1\\ 1 & \phantom{-}0\end{smallmatrix}\right)$.
These generators commute, $[\bA_x,\bA_y]=0$, because they are block-diagonal in a
common basis with each block a scalar multiple of $\bJ$. Commutativity is exactly the
condition under which the relative property of one-dimensional rotary encodings
\citep{su2024roformer} survives in more than one dimension; with non-commuting
generators the Baker--Campbell--Hausdorff correction leaves a residual dependence on
absolute position and the property fails \citep{heo2024rope,schenck2025string}.

\paragraph{Why physical units make a learned scale portable.}
In the original formulation $\theta_c$ is dimensionless, because positions are integer
sequence indices. Here $\lVert\btheta_c\rVert$ carries units of radians per
micrometre, so $2\pi/\lVert\btheta_c\rVert$ is a physical wavelength. The consequence
is that a length scale learned on one platform denotes the same relationship on
another. A rule of the form ``attend to my six nearest neighbours'' reaches roughly
$25\,\mu$m on a single-cell imaging panel and roughly $250\,\mu$m on a $100\,\mu$m
spot array, and so means something different on each; ``attend at $30\,\mu$m'' means
the same thing everywhere.

\section{Spatially informed gene ranking}
\label{app:sir}
For each cell, Tessera selects $K=500$ genes using
\emph{spatially informed ranking} (SIR), which
combines cell-specific expression relevance with
tissue-level spatial informativeness.
We select the top $K=500$ genes according to
$r_{j,i}$ for each cell. Thus, each cell receives
an adaptive gene set that reflects both its own
expression profile and the spatial organization
of gene expression across the surrounding tissue.
 Let $X\in\mathbb{R}^{M\times G}$ denote the gene-expression matrix
for a tissue section, where $M$ is the number of cells and $G$ the number of genes.

We first compute a cell-specific TF--IDF score. For gene $g_i$ in cell $s_j$,
\begin{equation}
\mathrm{TF}(g_i,s_j)
=
\frac{\log(1+X_{j,i})}
{\sum_{k=1}^{G}\log(1+X_{j,k})+\epsilon},
\end{equation}
and
\begin{equation}
\mathrm{IDF}(g_i)
=
\log\left(
\frac{1+M}
{1+\sum_{j=1}^{M}\mathbf{1}[X_{j,i}>0]}
\right).
\end{equation}
The resulting TF--IDF weight
\begin{equation}
w_{j,i}
=
\mathrm{TF}(g_i,s_j)\,\mathrm{IDF}(g_i)
\end{equation}
prioritizes genes that are prominent in a given cell while downweighting genes that
are broadly expressed across the section.

To incorporate tissue-level spatial organization, we additionally compute Moran's
$I$ \citep{moran1950notes} for each gene over the spatial coordinates of the section
using \textsc{Squidpy} \citep{palla2022squidpy}. Because TF--IDF and Moran's $I$
have different numerical scales, SIR combines their ranks rather than their raw
values. Let $\rho^{\mathrm{tf}}_{j,i}$ denote the TF--IDF rank of gene $i$ in cell
$j$, and $\rho^{\mathrm{sp}}_i$ its Moran's-$I$ rank across the section. We define
\begin{equation}
r_{j,i}
=
\frac{1-\lambda}{\kappa+\rho^{\mathrm{tf}}_{j,i}}
+
\frac{\lambda}{\kappa+\rho^{\mathrm{sp}}_i},
\qquad
\kappa=60,\quad \lambda=0.6,
\label{eq:sir}
\end{equation}
and select the top $K=500$ genes according to $r_{j,i}$ for each cell. Thus, each
cell receives an adaptive gene set that reflects both its own expression profile and
the spatial organization of gene expression across the surrounding tissue.

\section{Learning objective in full}
\label{app:objective}

\paragraph{Masking.}
Let $\mathcal{W}$ be a window of $N$ cells with coordinates $p_i \in \mathbb{R}^2$
in $\mu$m and raw integer counts $x_i \in \mathbb{Z}_{\geq 0}^{K}$ over the $K$ gene
tokens the cell carries. We draw a held-out set $\mathcal{M} \subset \mathcal{W}$ of
$\lceil \rho N \rceil$ cells uniformly at random, with $\rho = 0.10$. Masking is
applied to whole cells, never to individual genes or individual values. Writing
$z_i$ for the embedding produced by the frozen single-cell encoder, the masked
cell's token is replaced \emph{after} the input projection, so that no quantity
derived from its counts reaches the first attention layer:
\begin{equation}
h_i^{(0)} =
\begin{cases}
m, & i \in \mathcal{M},\\[2pt]
\mathrm{LN}\!\left(W_{\mathrm{in}} z_i\right), & \text{otherwise,}
\end{cases}
\end{equation}
where $m \in \mathbb{R}^{d}$ is a learned mask embedding. Coordinates are retained
for every cell, masked or not; position is never held out. Because cell embeddings
are precomputed and the encoder is bypassed during training, this ordering is
enforced by construction rather than by convention.

\paragraph{Library-size offset.}
The negative binomial mean is offset by a library size $\ell_i$. A masked cell's own
sequencing depth is itself part of what is held out, so using it would leak the
quantity being predicted; we replace it by the median depth of the visible cells in
the same window:
\begin{equation}
\ell_i =
\begin{cases}
\operatorname{median}\big\{\, \textstyle\sum_k x_{jk} \;:\; j \in \mathcal{W} \setminus \mathcal{M} \,\big\}, & i \in \mathcal{M},\\[4pt]
\textstyle\sum_k x_{ik}, & \text{otherwise.}
\end{cases}
\end{equation}

\paragraph{Decoder.}
The geometry-aware representation $h_i^{(L)}$ is decoded by two linear heads into
the parameters of a gene-wise negative binomial over the decoder vocabulary of size
$G$:
\begin{align}
\log \mu_{ig} &= \log \ell_i \;-\; \log G \;+\; e_g^{\top} h_i^{(L)} + b_g, \\
\log \theta_{ig} &= \operatorname{clip}\!\left(w_g^{\top} h_i^{(L)} + c_g,\; -\tau,\; \tau\right),
\qquad \tau = 10 .
\end{align}
The $-\log G$ term initialises the decoder at a uniform expression profile, and
$\log \mu$ is clipped above at $30$ for numerical stability.

\paragraph{Loss.}
Writing $g_{ik}$ for the vocabulary index of the $k$-th gene token of cell $i$, the
objective is the negative binomial negative log-likelihood on the raw counts of the
held-out cells only:
\begin{equation}
\mathcal{L} \;=\; \frac{1}{|\mathcal{M}|\,K}
\sum_{i \in \mathcal{M}} \sum_{k=1}^{K}
-\log \mathrm{NB}\!\left(x_{ik} \,\middle|\, \mu_{i g_{ik}},\, \theta_{i g_{ik}}\right),
\end{equation}
with
\begin{multline}
-\log \mathrm{NB}(x \mid \mu, \theta) = -\Big[
\log \Gamma(x + \theta) - \log \Gamma(\theta) - \log \Gamma(x + 1) \\
+\; \theta\big(\log \theta - \log(\theta + \mu)\big)
\;+\; x\big(\log \mu - \log(\theta + \mu)\big) \Big].
\end{multline}
The loss is gathered onto the $K$ gene tokens each cell actually carries, so genes
outside a slide's panel contribute no gradient, and the single-cell encoder is
frozen throughout.

\section{Experiments}
\label{app:experiments}

All results are evaluated on held-out whole sections. We use the spatially resolved subset of SpatialCorpus-110M \citep{schaar2025nicheformer}, comprising $43$ sections and $20{,}068{,}822$ cells, split $30/4/9$ by section. The nine held-out sections span two imaging technologies and seven tissues and are used only for final evaluation. Our primary statistic is the paired held-out imputation Pearson correlation, computed as a section-wise difference over $n=9$ sections ($\mathrm{df}=8$, two-sided critical value $2.306$) and bootstrapped over sections rather than cells. All thresholds reported below were preregistered before the corresponding results were observed; the complete register, including failed criteria, is given in Appendix~\ref{app:register}.
\paragraph{Models.}
\Ody{} combines a frozen scGPT whole-human backbone
\citep{cui2024scgpt} with a trainable spatial attention
module. For each cell, the backbone encodes $K=500$
gene tokens selected through spatially informed ranking
into a single $512$-dimensional cell representation.
The spatial attention module consists of $L=6$ layers,
$8$ attention heads, and a hidden dimension of
$d_{\mathrm{model}}=512$, comprising approximately
$19$M trainable parameters. It operates on windows
of $N=256$ spatially contiguous cells with a
$d_{\max}=200\,\mu\mathrm{m}$ distance cutoff.

The positional-encoding comparison in
\S\ref{sec:pe} evaluates four spatial encoding
strategies within this shared architecture:
absolute sinusoidal embeddings of $(x,y)$,
RoPE, PoPE~\citep{gopalakrishnan2026pope},
and an additive radial distance bias parameterized
using $32$ Gaussian basis functions
\citep{zhou2023unimol,zhao2025stofm}.
We additionally include a geometry-free baseline
without positional information. All five
configurations share the same pretrained backbone,
spatial attention architecture, training windows,
masking strategy, and optimization protocol.
They differ in their spatial encoding mechanisms,
including the query--key parameterization
associated with PoPE.

We also evaluate five non-spatial predictors
under matched evaluation conditions.
Four non-learned baselines comprise
\texttt{mean\_profile},
\texttt{neighbour\_average},
PCA on highly variable genes, and PCA on
neighbourhood-averaged expression.
The fifth baseline,
\texttt{scgpt\_no\_aggregator},
uses the frozen scGPT backbone embeddings
directly to assess the contribution of
spatial contextualization.

For the comparison with existing spatial
foundation models in \S\ref{sec:benchmark},
we use the released pretrained weights and
original preprocessing pipelines of Nicheformer
\citep{schaar2025nicheformer} and scGPT-spatial
\citep{wang2025scgptspatial}.
All three models produce $512$-dimensional
per-cell embeddings, verified at runtime,
so no additional dimensionality matching
is required.

\paragraph{Training details.}
All models are trained with the cross-cell cloze objective:
a fraction $\rho = 0.1$ of cells in each window is masked, and the negative-binomial
loss is computed on masked cells only over a fixed decoder vocabulary of $1{,}245$
genes. 
For each cell, we select the top $K=500$ genes according to the
spatially informed ranking score in Eq.~\eqref{eq:sir}, with
$\lambda=0.6$ and $\kappa=60$ fixed on development sections before
held-out evaluation. The selected genes are encoded by the frozen
scGPT backbone, and the resulting cell embeddings are cached for
spatial model training and evaluation.. We optimise with AdamW
($\beta = (0.9, 0.95)$, weight decay $0.05$) at a peak learning rate of
$3\times10^{-4}$, linearly warmed up over $500$ steps and then cosine-annealed to
zero, with gradients clipped to a global norm of $1.0$, in bf16, at a batch of eight
windows ($2{,}048$ cells) per step and one accelerator per arm with arms run
concurrently. The aggregator and the decoder share that learning rate; the only
parameters on a separate schedule are the rotary and PoPE frequency grids, which
carry a $0.1\times$ multiplier and so train at $3\times10^{-5}$, a choice we found
necessary to prevent an early collapse of attention entropy. Rotary frequencies are
initialised on a geometric grid spanning $10\,\mu$m to $2\,$mm with bearings uniform
on $[0,\pi)$. The main results use $50{,}000$ steps; the positional-encoding
comparison of \S\ref{sec:pe} and the six-seed replication use $18{,}000$ steps on a
narrower $948$-gene decoder vocabulary. The two budgets are identical in every
optimisation setting above and differ only in the step count and hence in the cosine
decay horizon, and their absolute values are not comparable, so no comparison in this
paper crosses that boundary. Within each budget every arm is byte-identical in
configuration and seed.
\paragraph{Datasets and Evaluation details}

Training masks whole cells. For a fraction $\rho = 0.1$ of each window we replace the
cell token by a learned mask embedding while retaining the coordinate, so a masked
cell's output state $\bz_i$ is a function of the cells around it, their positions and
its own position, and of nothing else about it. A negative-binomial head decodes
counts, with $\log \hat{\mu}_{ig} = [\mathrm{MLP}(\bz_i)]_g + \log \varsigma_i$ for
library size $\varsigma_i$, and the loss
$-\sum_{i \in \Mask} \sum_g \log \mathrm{NB}(x_{ig} \mid \hat{\mu}_{ig}, \phi_g)$ is
computed on masked cells only. Reconstructing a cell from its neighbourhood is an
established objective family
\citep{fischer2023ncem,wen2023cellplm,zhao2025stofm}. The task cannot be solved without geometry, since position is the only
cell-specific information retained about a masked cell. Masking is applied after the frozen backbone and input projection:
each masked cell's projected embedding is replaced by a learned mask
token before entering the spatial attention module, while its coordinates
are retained (Fig.~\ref{fig:arch}; Appendix~\ref{app:objective}).
Thus, the masked cell's own expression-derived embedding is not
available to the spatial attention module.

Measuring what a component buys usually means comparing a model against a
re-implementation of a baseline, which mixes the mechanism under test with everything
else two codebases differ in. We remove that confound structurally: the output
projection of every inter-cell attention block is zero-initialised, so at step $0$ the
aggregator is the identity map and a model with a positional encoding emits outputs
identical to one without. The observed deviation is exactly $0$ on real checkpoints
rather than merely within tolerance, and in the positional-encoding comparison of
\S\ref{sec:pe}, all five configurations,
including the geometry-free baseline, begin
training at an identical loss to numerical
precision. Arms share the
seed, the window stream, the mask draw, the split and the schedule, and differ only in
the encoding under test. The baseline is not an approximation of our model without
geometry; it is our model without geometry.

\section{Spatially informed gene ranking improves spatial representations}
\label{app:ranking}

We further evaluate whether spatially informed gene ranking (SIR) improves
representations for a downstream spatial transcriptomics task. On the HER2 cohort, we
cluster cells using an identical Scanpy pipeline while varying only the gene-selection
strategy: HVG, SVG, or SIR. HVG and SVG assign a fixed gene set across a tissue
section, whereas SIR constructs a cell-specific ranking by combining local expression
informativeness with tissue-level spatial structure through TF--IDF and Moran's $I$.
SIR consistently yields more discriminative spatial representations, achieving the
highest kNN purity ($0.8149$), ARI ($0.5107$), and Silhouette score ($-0.0007$)
(Table~\ref{tab:gene_ranking}). These results suggest that jointly modeling
cell-specific and tissue-level spatial signals selects genes that better preserve
biologically relevant heterogeneity for downstream spatial analysis.
\begin{table}[t]
\centering
\small
\caption{\textbf{Effect of gene selection on spatial representation quality.}
Clustering of the HER2 cohort using the same downstream pipeline while varying only
the input gene-selection strategy. Higher is better for all metrics.}
\label{tab:gene_ranking}
\begin{tabular}{lccc}
\toprule
Method & kNN Purity $\uparrow$ & ARI $\uparrow$ & Silhouette $\uparrow$ \\
\midrule
HVG & 0.6236 & 0.2957 & -0.0323 \\
SVG & 0.7666 & 0.2928 & -0.0280 \\
\textbf{SIR (Ours)} & \textbf{0.8149} & \textbf{0.5107} & \textbf{-0.0007} \\
\bottomrule
\end{tabular}
\end{table}

\section{Positional encoding comparison}
\label{app:pe}

Table~\ref{tab:pe} gives the full four-arm comparison summarised in
\S\ref{sec:pe}. All four arms share the architecture, the seed, the cached cell
tokens, the window stream, the mask draw and the schedule, and differ only in the
positional term of the attention logit. Because every added component is
zero-initialised, the four begin training at an identical loss to sixteen digits, so
each $\Delta$ is attributable to the encoding rather than to any difference in
initialisation or data order. Figure~\ref{fig:f15} shows what the encodings do rather
than how well they score, and the per-section values in
Table~\ref{tab:persection} confirm that the ordering is not driven by any one
section.
\begin{table}[t]
\centering
\scriptsize
\caption{
\textbf{Effect of positional encoding on spatial gene reconstruction.}
Four spatial encoding strategies are compared against a
geometry-free baseline under a matched training protocol.
Pearson correlation is averaged across nine held-out sections;
$\Delta$ and $t$ report the improvement and paired test statistic
relative to the baseline ($\mathrm{df}=8$).
\textsc{Mass} denotes the fraction of attention within
$25\,\mu$m, and \textsc{Rose CV} measures directional
anisotropy (higher is more directional).
Relative encodings substantially improve reconstruction,
with RoPE achieving the highest correlation, spatial locality,
and directional anisotropy.
}
\label{tab:pe}
\begin{tabular}{lccccc}
\toprule
Encoding & Pearson $r$ & $\Delta r$ & $t$ &
Mass ${<}25\,\mu$m & Rose CV \\
\midrule
No positional information
& $0.2575$ & --- & --- & $0.068$ & $0.008$ \\

Absolute sinusoidal
& $0.2580$ & $+0.0005$ & $1.18$ & $0.067$ & $0.007$ \\

PoPE \citep{gopalakrishnan2026pope}
& $0.4235$ & $+0.1660$ & $9.04$ & $0.759$ & $0.207$ \\

Additive radial distance bias
& $0.4176$ & $+0.1602$ & $9.07$ & $0.468$ & $0.050$ \\

\midrule
\textbf{RoPE (Ours)}
& $\mathbf{0.4311}$ & $\mathbf{+0.1736}$ &
$\mathbf{9.44}$ & $\mathbf{0.826}$ & $\mathbf{0.490}$ \\
\bottomrule
\end{tabular}
\end{table}
Table~\ref{tab:invariance} reports what changes under a rigid motion. This appendix
states why the rotation entry is a design property rather than a defect.

Because $\psi_c(\bp) = \btheta_c^{\!\top}\bp$ and hence
$\psi_c(Q\bp) = (Q^{\!\top}\btheta_c)^{\!\top}\bp$ for any rotation $Q$, rotating
the tissue is exactly equivalent to rotating the frequency field. We verified this:
rotating the data alone changes the logits by $2.874$, while rotating the data and
setting $\btheta \leftarrow Q\btheta$ changes them by $3.6\times10^{-14}$ in
float64. The encoding is therefore rotation-\emph{equivariant} rather than
rotation-invariant, and the direction it reads is expressed in the laboratory frame.

\begin{table}[h]
\centering
\scriptsize
\caption{\textbf{What each encoding can see.} ``Uses direction'' means the attention
logit depends on the bearing of $\bD_{ij}$ and not only on its magnitude.
``Equivariant'' means a rotation of the tissue is exactly equivalent to a rotation of
the frequency field, so the representation transforms predictably rather than staying
fixed.}
\begin{tabular}{lcccc}
\toprule
Encoding & translation & rotation & reflection & uses direction \\
\midrule
no positional information & invariant & invariant & invariant & no \\
absolute sinusoidal & no & no & no & yes, and absolute position \\
PoPE & invariant, exactly & equivariant & no & yes \\
radial bias $b_{\mathrm{rad}}(d)$ & invariant & invariant & invariant & no \\
angular bias in a local frame & invariant & invariant & invariant & yes, co-rotating frame \\
\midrule
\textbf{RoPE (this work)} & invariant, exactly & equivariant & no & yes, in the lab frame \\
\bottomrule
\end{tabular}
\end{table}


Two entries deserve comment. An angular bias measured against a data-derived local axis uses direction while remaining rotation-invariant, because the frame
co-rotates with the tissue; direction and rotation invariance are therefore not
strictly incompatible in general, though they are for any encoding that reads
direction in the laboratory frame, as ours does. Conversely, a full pairwise
distance matrix determines a configuration up to rigid motion \emph{and reflection}, so an SE(2) architecture built on one is more expressive than the per-pair bias; what it cannot do at any depth or capacity is distinguish a configuration from its mirror image. Our encoding does, with a reflection drift of $30.90$ against a live signal of $26.38$.

\begin{figure}[t]
\centering
\includegraphics[width=0.85\textwidth]{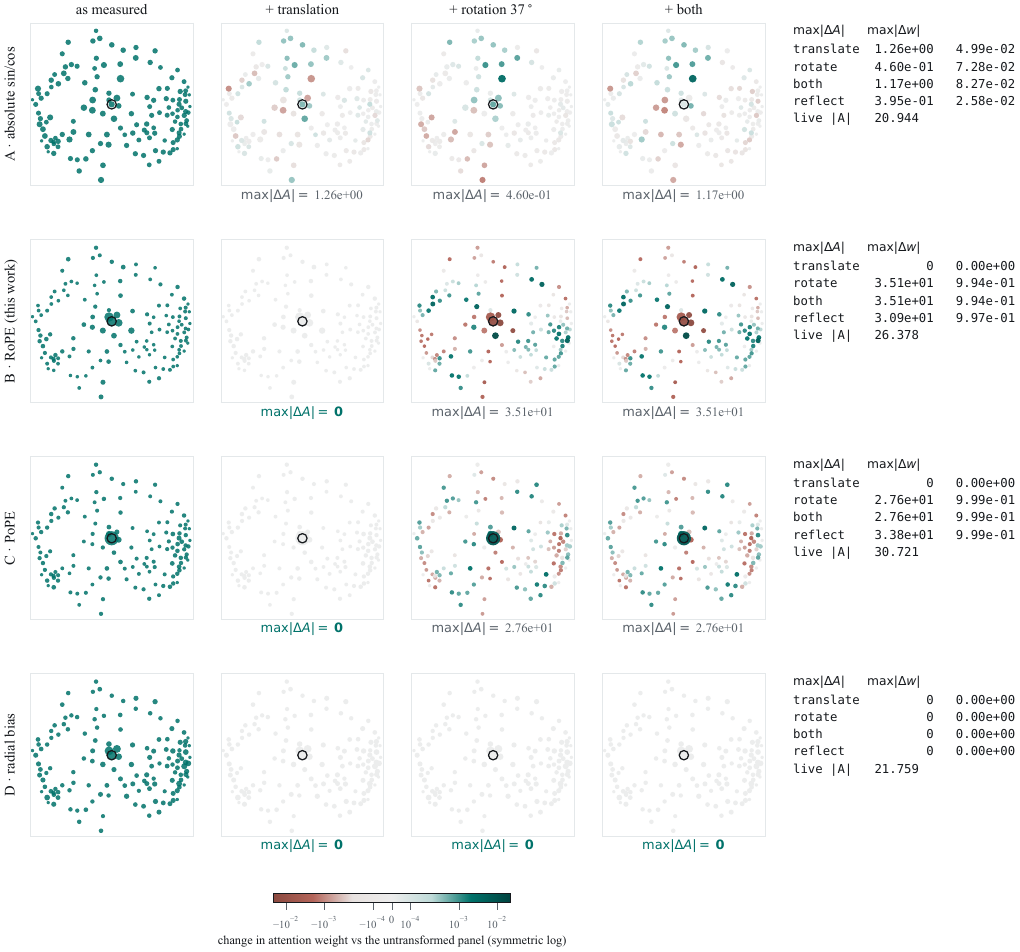}
\caption{\textbf{Attention under spatial transformations.}
Attention patterns for a fixed query cell under translation, rotation
by $37^\circ$, and both combined. RoPE and PoPE are translation-invariant
but rotation-sensitive; the radial bias is invariant to both, while
absolute sinusoidal encoding is sensitive to both. Tables report maximum
absolute changes in attention logits and weights, evaluated in float64.}
\label{fig:f16}
\end{figure}

\begin{table}[t]
\centering
\scriptsize
\caption{\textbf{Sensitivity to rigid motion.} Largest absolute change in the
attention logit when a held-out window is translated
($12.3\times8.9\,$mm), rotated ($37^\circ$ about its centroid) or reflected,
over eight windows. \textsc{Live} $|A|$ is the largest untransformed logit, so no row
can appear invariant by producing no signal. Zeros are exact in float64 and asserted:
for a relative encoding the translation entry is an identity, so a non-zero raises an
error rather than being reported.}
\label{tab:invariance}
\begin{tabular}{lcccc}
\toprule
Encoding & translation & rotation $37^\circ$ & reflection & live $|A|$ \\
\midrule
no positional information & $0$ & $0$ & $0$ & $26.56$ \\
absolute sinusoidal & $1.256$ & $0.460$ & $0.395$ & $20.94$ \\
PoPE & $0$ & $27.63$ & $33.79$ & $30.72$ \\
radial distance bias & $0$ & $0$ & $0$ & $21.76$ \\
\midrule
\textbf{RoPE (this work)} & $\mathbf{0}$ & $35.05$ & $30.90$ & $26.38$ \\
\bottomrule
\end{tabular}
\end{table}

\section{Extrapolation beyond the training window}
\label{app:extrap}
We evaluate zero-shot extrapolation by increasing the window size
from $N=256$ to $N=4096$ cells without retraining or rescaling
positional frequencies. In the primary experiment, the attention
cutoff scales with the measured window radius, using the same rule
for all four positional encodings. This expands both the number
of cells and the physical range available to attention.
Under this protocol, held-out reconstruction loss increases by
$1.6\%$ for RoPE and $2.0\%$ for PoPE, compared with $23\%$
for the radial distance bias (Fig.~\ref{fig:f17}).

As a control, we increase the window size while holding the
attention cutoff fixed at $200\,\mu$m. The fraction of cell pairs
within the cutoff decreases from $0.919$ to $0.186$, while
reconstruction loss changes only modestly. This supports the
interpretation that the differences in Fig.~\ref{fig:f17}
primarily reflect extrapolation to a larger physical attention
range rather than an increase in cell count alone.

\section{Benchmark protocol and full results}
\label{app:benchmark}

The task list, the probe protocol, the validation gate and its threshold, the fit
budget and the cell-intersection rule were all written before any baseline was
staged. Two tasks were declared unevaluable in advance and are reported unevaluated
rather than dropped.

The cohort is $39$ sections and $18{,}319{,}942$ cells: $30$ training sections on
which probes are fitted, and nine held-out sections on which every number is scored.
Cell ordering was verified against the cached representation on all $39$ sections
for every model. No model was scored until it passed a strict weight load with zero
missing and zero unexpected tensors, an embedding rank equal to the model width,
exact determinism on repeated embedding, and a gene-shuffle negative control;
observed degradation fractions under the shuffle were $0.733$ for \Ody, $0.818$ for
Nicheformer and $0.856$ for scGPT-spatial, against a pre-registered bar of $0.50$.

Every model is scored on the same $4{,}986{,}465$ cells, an intersection over all
models. The $43{,}850$ excluded cells are all ones scGPT-spatial cannot embed,
because their counts over the matched panel are entirely zero and a frozen per-cell
encoder that keeps only non-zero genes has nothing to tokenise. Such cells carry a
missing row rather than a zero row, since a zero row is a fabricated embedding at
the origin, and \Ody{} was denied those cells too.

Niche composition and cell-type annotation require cell-type labels, which the nine
held-out sections do not carry. Scoring them on the three labelled training sections
was refused, since \Ody{} trained on those sections and the baselines did not.
Neighbourhood density is the closest label-free substitute and is what we report. A
third registered task, imputation through the shared linear probe, placed all three
models below a constant predictor: one ridge cannot map a $512$-dimensional
embedding onto a $1245$-gene count vector, and it fails identically for all three,
so we report a failed readout rather than a result and did not substitute a
statistic that works better.

\begin{figure}[h]
\centering
\includegraphics[width=0.60\textwidth]{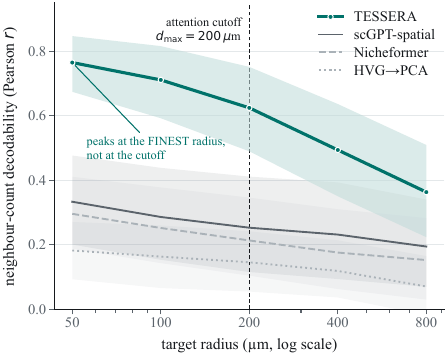}
\caption{Neighbour-count decodability against target radius; bands are $95\%$
bootstrap intervals over nine sections and the dashed line marks the $200\,\mu$m
attention cutoff. Had the model learned its own cutoff the curve would peak there;
it does not, decreasing monotonically and reaching its maximum at the finest radius
tested. The registered criterion holds at $50/100/200\,\mu$m and fails at
$400/800\,\mu$m, and the verdict is unanimous across six training seeds.}
\label{fig:radius}
\end{figure}

\begin{table}[h]
\centering
\scriptsize
\caption{Neighbour-count decodability by target radius. Left: a single checkpoint at
$50{,}000$ steps with the best baseline for reference. Right: six seeds at
$18{,}000$ steps, mean $\pm$ standard deviation, with the per-seed verdict under the
registered criterion.}
\begin{tabular}{lccccc}
\toprule
Radius & \Ody{} ($50$k) & best baseline & six-seed mean $\pm$ SD & seeds above & verdict \\
\midrule
$50\,\mu$m & $0.7651$ & $0.3332$ & $0.7234 \pm 0.0241$ & $6/6$ & \textbf{holds $6/6$} \\
$100\,\mu$m & $0.7113$ & $0.2858$ & $0.6698 \pm 0.0229$ & $6/6$ & \textbf{holds $6/6$} \\
$200\,\mu$m & $0.6246$ & $0.2526$ & $0.5894 \pm 0.0197$ & $6/6$ & \textbf{holds $6/6$} \\
$400\,\mu$m & $0.4938$ & $0.2312$ & $0.4762 \pm 0.0158$ & $6/6$ & fails $0/6$ \\
$800\,\mu$m & $0.3630$ & $0.1941$ & $0.3708 \pm 0.0222$ & $6/6$ & fails $0/6$ \\
\bottomrule
\end{tabular}
\end{table}

The $400$ and $800\,\mu$m entries are interval failures rather than ordering
failures: the sign holds on all six seeds at every radius, and even at $800\,\mu$m
the weakest seed reaches $0.3417$ against the best baseline's $0.1941$. Two caveats
belong with this table. The seeds are $18{,}000$-step checkpoints against a
$50{,}000$-step headline, so the spread measures training-seed variability at the
smaller budget; and the baselines are released frozen artefacts with no seed, so the
single-seed limitation is removed on our side only.

On technology decodability, a probe recovers the generating platform from our
embedding at $\mathrm{AUC} = 0.986$, third of four behind HVG$\to$PCA at $0.457$ and
scGPT-spatial at $0.866$ and ahead only of Nicheformer at $0.99998$. Compressing
every embedding to $32$ dimensions does not remove it --- ours rises slightly to
$0.993$ --- so the platform signal lives in the leading variance directions rather
than spread thin. Notably, being platform-decodable did not prevent the biology from
transferring: Nicheformer is the most decodable model of the four and also the one
that collapses to $0.004$ across platforms.

\section{Per-section results}
\label{app:register}

\begin{table}[h]
\centering
\scriptsize
\caption{\textbf{The encoding comparison, all nine held-out sections}, no selection.
Values are held-out imputation Pearson at $18{,}000$ steps, seed $0$.}
\label{tab:persection}
\begin{tabular}{lccccc}
\toprule
Section & none & absolute & PoPE & radial & \textbf{RoPE} \\
\midrule
Xenium breast, invasive lobular & $0.2366$ & $0.2379$ & $0.3592$ & $0.3481$ & $\mathbf{0.3636}$ \\
Xenium colon & $0.3372$ & $0.3370$ & $0.5210$ & $0.5194$ & $\mathbf{0.5232}$ \\
Xenium lung, non-diseased & $0.1237$ & $0.1231$ & $0.2348$ & $0.2332$ & $\mathbf{0.2440}$ \\
Xenium melanoma & $0.3338$ & $0.3371$ & $0.5362$ & $0.5324$ & $\mathbf{0.5453}$ \\
MERSCOPE breast, patient 1 & $0.2273$ & $0.2276$ & $0.4765$ & $0.4713$ & $\mathbf{0.4868}$ \\
MERSCOPE lung, patient 2 & $0.3265$ & $0.3264$ & $0.5031$ & $0.4774$ & $\mathbf{0.5086}$ \\
MERSCOPE ovary, patient 2 & $0.2583$ & $0.2597$ & $0.4615$ & $0.4566$ & $\mathbf{0.4745}$ \\
MERSCOPE uterus, patient 1 & $0.2923$ & $0.2922$ & $0.4678$ & $0.4621$ & $\mathbf{0.4705}$ \\
MERFISH liver, HCC & $0.1814$ & $0.1808$ & $0.2513$ & $0.2583$ & $\mathbf{0.2630}$ \\
\bottomrule
\end{tabular}
\end{table}

Our margin over the no-position arm ranges from $+0.082$ to $+0.260$ and is positive on every
section, whereas the absolute encoding's ranges from $-0.0006$ to $+0.0033$.

\section{Limitations} \label{app:limits}


 The backbone is frozen, so \Ody{} is an aggregator over a single-cell embedding
 rather than an end-to-end spatial foundation model, and its ceiling is that of the
 embedding it consumes. Unfreezing it did not change the result, but four factors
change together between those two arms, so we register and report that comparison as
 descriptive rather than as an isolated measurement.





\end{document}